\documentclass[12pt]{iopart}
\usepackage[dvips]{graphics, color}
\usepackage{iopams,wrapfig}
\usepackage{xspace}
\usepackage{subfig}

\begin{document}

\title[$\gamma$-jet measurements]{$\gamma-$hadron azimuthal correlations in STAR}

\author{A M Hamed for the STAR Collaboration}

\address{Texas A\&M University, College Station, TX 77843, USA}
\ead{ahamed@tamu.edu}
\begin{abstract}
Azimuthal correlations of direct photons at high transverse energy (8 $<$ E$_T$ $<$ 16 GeV) 
with away-side hadrons of transverse momentum (3 $<$ p$_T$ $<$ 6 GeV/c) have been measured over a broad range of centrality for 
$Au+Au$ collisions and $p+p$ collisions at $\sqrt{s_{NN}}$ = 200 GeV. The presented results are the  
first measurements at RHIC for $\gamma$-hadron azimuthal correlations in $Au+Au$ collisions.

\end{abstract}
\section{Introduction}
One of the most exciting results at RHIC is the suppression of hadrons in central $Au+Au$ collisions, 
compared to $p+p$ collisions [1] and to cold nuclear matter [2] at high p$_{T}$. 
This suppression is well described by very different pQCD-based energy loss models in the light flavor sector [3]. However, the observation 
of a similar level of suppression for the heavy flavor [4], and the unexpected smaller supression of baryons compared to mesons  
at intermediate to high p$_{T}$ [5], cannot be easily reconciled in these pQCD-based energy loss models that incorporate 
gluon radiation as the main source of in-medium energy loss. In fact the applicability of pQCD in describing the parton-matter interaction has been increasingly challenged 
by the strongly coupled nature of the produced matter at RHIC. As a result there is no single commonly accepted calculation of 
the underlying physics to describe in-medium energy loss for different quark generations as well as for the gluon.
 
On the other hand, experimental observables based on single-particle spectra are not sensitive enough to discriminate between the different energy loss mechanisms, indicating the need for additional experimental observables in order to better constrain 
the energy loss mechanism. Azimuthal correlation measurements of di-hadron and $\gamma$-hadron provide a complementary way to quantify the energy loss, and its dependence on  
path length, and the initial parton energy. In particular, $\gamma$-hadron is an ideal probe [6] for the dependence on the initial parton energy 
and possibly the color-factor. 

\section{Data and Analysis}
Using a level-2 high-$p_T$ tower trigger to tag $\gamma$-jet events, in 2007 the STAR experiment collected an integrated
luminosity of 535 $\mu {b}^{-1}$ of $Au+Au$ collisions at $\sqrt{s_{NN}}$ = 200 GeV. The BEMC has a full azimuthal coverage and
pseudorapidity coverage $\mid\eta\mid$ $\leq$ 1.0. As a reference measurement we use $p+p$ data at $\sqrt{s_{NN}}$ = 200 GeV
taken in 2006 with integrated luminosity of 11 $\mathrm{pb}^{-1}$. The Time Projection Chamber (TPC) was used
to detect charged particle tracks and measure their momenta. For this analysis, events with at least
one cluster with $E_T >$ 8~GeV were selected.

A crucial step of the analysis is to discriminate between showers of direct $\gamma$ and two close $\gamma$'s from a
high-p$_{T}$ $\pi^{0}$ symmetric decay. At p$_T$ $\sim$ 8 GeV/c the angular separation between the two photons resulting from a $\pi^{0}$ ``symmetric decay"
at the BEMC face is typically smaller than the tower size ($\Delta\eta=0.05,\Delta\phi=0.05$); but 
a $\pi^{0}$ shower is generally broader than a single $\gamma$ shower. The Barrel Shower Maximum Detector (BSMD), which resides at $\sim$ 5X$_{0}$ inside the calorimeter towers, is well-suited for 
$(2\gamma$)/$(1\gamma)$ separation up to p$_T$ $\sim$ 26 GeV/c due to its fine segmentation ($\Delta\eta\approx 0.007,\Delta\phi\approx 0.007$). 
In this analysis the $\pi^{0}$/$\gamma$ discrimination was carried out by making cuts on the shower shape as measured by the BSMD, where the $\pi^{0}$ identification 
cut is adjusted in order to obtain very pure sample of $\pi^{0}$ and a sample rich in direct $\gamma$ ($\gamma_{rich}$).
The near- and away-side yields, Y$^{n}$ and Y$^{a}$, of associated particles per trigger are extracted by integrating the $\rm dN/{\rm d}(\Delta\phi)$  
distributions in $\mid\Delta\phi\mid$ $\leq$  0.63 and $\mid\Delta\phi -\pi\mid$  $\leq$  0.63 respectively. 
The yield is corrected for the tracking efficiency of associated charged particles as a function of multiplicity.

The shower shape cuts used to select a sample of direct-photon-``rich" triggers reject most of the $\pi^{0}$'s, but do not reject 
photons from highly asymmetric $\pi^{0}$ decays, $\eta$'s, and fragmentation photons. 
All of these sources of background get removed as follows (Eq 1), but only within the systematic
uncertainty on the assumption that their correlations are similar to those for $\pi^{0}$'s.
Assuming zero near-side yield for direct photon triggers and a very pure sample of $\pi^{0}$, the away-side yield of hadrons correlated with the direct photon is extracted as
\begin{eqnarray}
Y_{\gamma_{direct}+h}=\frac{Y^{a}_{\gamma_{rich}+h}-R Y^{a}_{\pi^{0}+h}}{1-R},\label{eq2}
\hspace{0.5cm}                            
where   \hspace{0.5cm}   R=\frac{Y^{n}_{\gamma_{rich}+h}}{Y^{n}_{\pi^{0}+h}}.
\end{eqnarray}

\begin{figure}
\begin{center}
\begin{tabular}{cc}
   \resizebox{140mm}{!}{\includegraphics{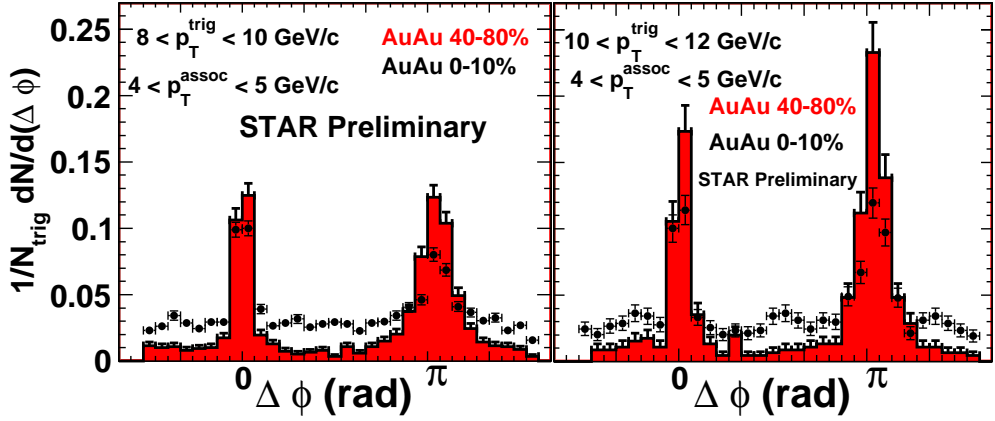}} 
        \end{tabular}
    \caption{Azimuthal correlation histograms of high p$_{T}^{trig}$ inclusive photons 
    with associated hadrons for 40-80$\%$ and 0-10$\%$ $Au+Au$
    collisions.}
      \end{center}
\end{figure}

\section{Results}
Figure 1 shows the azimuthal correlation for inclusive photon triggers for the most peripheral and central bins. 
Parton energy loss in the medium causes the away-side to be increasingly suppressed with centrality as it was previously reported [2,7].
The suppression of the near-side yield with centrality, which has not been observed in the charged hadron 
azimuthal correlation, is consistent with an increase of the $\gamma$/$\pi^{0}$ ratio with centrality at high E$_{T}^{trig}$. 
\begin{figure}
  \begin{center}
    \begin{tabular}{cc}
           \resizebox{100mm}{!}{\includegraphics{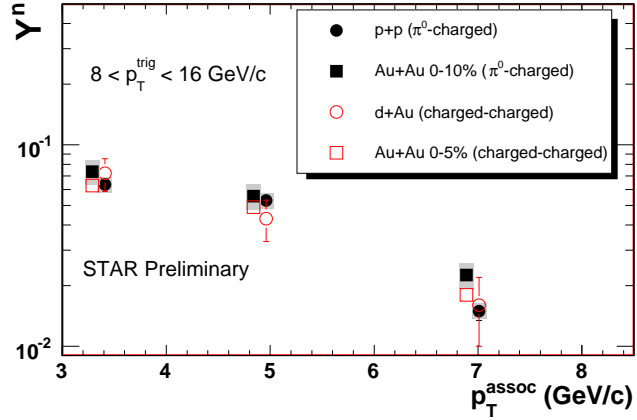}} 
    \end{tabular}
    \caption{$p_{T}^{assoc}$ dependence of $\pi^{0}$-ch and $ch-ch$ [7] near-side yields.}
         \end{center}
\end{figure} 
Figure 2 shows the $p_{T}^{assoc}$ dependence of the near-side associated yields for $\pi^{0}$-charged correlations ($\pi^{0}$-ch)
compared to measurements with charged hadrons only (ch-ch) [7]. A general agreement between both types of analysis is found, indicating that the
$\pi^{0}$ selection is relatively pure.
\begin{figure}
 \begin{center}
  \begin{tabular}{cc}
   \resizebox{160mm}{!}{\includegraphics{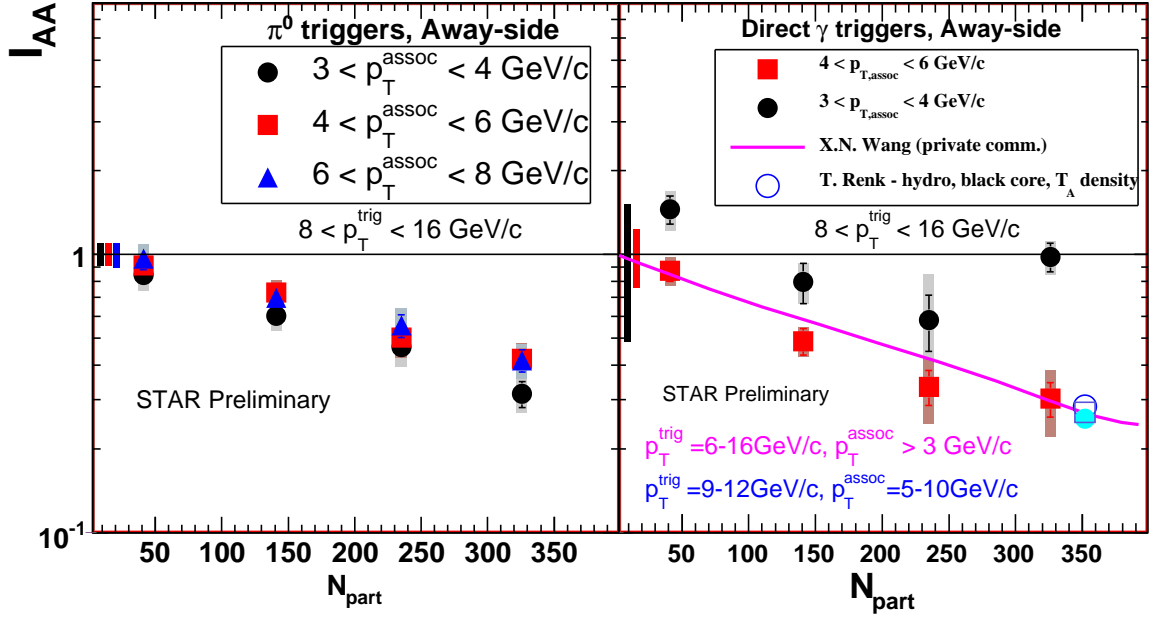}} 
     \end{tabular}
    \caption{(Right) I$_{AA}$ for $\pi^{0}$ triggers. (Left) I$_{AA}$ for direct $\gamma $ triggers (see text). Boxes on the left show the scale uncertainty due to $p+p$ measurements.}
    \end{center}
\end{figure}

In order to quantify the away-side suppression, we calculate the quantity I$_{AA}$, which is defined as the ratio of the integrated yield of 
the away-side associated particles per trigger particle in $Au+Au$ relative to $p+p$ collisions.
Figure 3 (left) and Fig. 3 (right) show the I$_{AA}$ for $\pi^{0}$ triggers and for direct $\gamma$ triggers respectively, 
as a function of centrality.
The ratio would be unity if there were no medium effect on the parton fragmentation; indeed the most
peripheral bin shows a ratio close to unity. The ratio decreases with centrality for more central events in a similar fashion for  
different ranges of p$_{T}^{assoc}$. Although $\pi^{0}$ results from higher parton energy and is surface-biased, 
the suppression in the $\gamma$-triggered away-side yield is similar to that of the $\pi^{0}$-triggered sample
within the measured momentum range. The statistical and systematic uncertainties on the $p+p$ measurement result in a rather large scale
uncertainty, which can be reduced with larger data samples that are expected in the near future.
The value of I$_{AA}$ for $\gamma$-hadron correlations in the most central events is found to be similar to the values observed for di-hadron
correlations and for single-particle suppression R$_{AA}$ and agrees well with theoretical calculations in which the energy loss is tuned to the
single- and di-hadron measurements [8,9]. 

In conclusion, a first measurement of $\gamma$-jet in $Au+Au$ collisions has been performed by the STAR experiment. 
Within the current uncertainty in scaling, the I$_{AA}$ agrees with the theoretical calculations.


\section*{References}
\begin{thebibliography}{9}
\bibitem{} S. Adler et al., {Phys. Rev. Lett.91 072303 (2003)}
\bibitem{} J. Adams et al., {Phys. Rev. Lett.91 072304 (2003)}
\bibitem{} A. Majumder, {J.Phys.G34 S377-388 (2007) and references therein} 
\bibitem{} B.I. Abelev et al., {Phys. Rev. Lett. 98 192301 (2007)}
\bibitem{} P. Fachini, {these proceedings}
\bibitem{} X.N.Wang, Z. Huang, and I. Sarcevic., {Phys. Rev. Lett. 77 (1996)}
\bibitem{} J. Adams et al., {Phys. Rev. Lett. 97 162301 (2006)}
\bibitem{} X. N. Wang and H. Zhang, {private communication}  
\bibitem{} T. Renk, {Phys. Rev. C74 034906 (2006).} 
\end{thebibliography}
\end{document}